\documentclass{article}

\usepackage{tabularx}

\usepackage{multirow}

\title{On the Fly Query Entity Decomposition Using Snippets}

\author{David J. Brenes\footnote{Simplelogica S.L.: david.brenes@simplelogica.net}, Daniel Gayo-Avello\footnote{ University of Oviedo: dani@uniovi.es} and Rodrigo Garcia\footnote{University of Oviedo: rodrigo@innova.uniovi.es}}

\begin{document}

\maketitle

\begin{abstract}

One of the most important issues in Information Retrieval is inferring the intents underlying users' queries. Thus, any tool to enrich or to better contextualized queries can proof extremely valuable. Entity extraction, provided it is done fast, can be one of such tools. Such techniques usually rely on a prior training phase involving large datasets. That training is costly, specially in environments which are increasingly moving towards real time scenarios where latency to retrieve fresh informacion should be minimal. In this paper an `on-the-fly' query decomposition method is proposed. It uses snippets which are mined by means of a naïve statistical algorithm. An initial evaluation of such a method is provided, in addition to a discussion on its applicability to different scenarios.

\end{abstract}

\section{Introduction}

One of the main purposes in Web Information Retrieval is helping users to fulfill their information needs. However, most of the time, the main problem is determining the users' intent. For instance, is s/he looking for a particular result? A comparison between some results? Does the input provide any clue about the goal or the intent behind the query? Has any other user issued the same query before? Nonetheless to say, interpreting the users' input is a complex task that can be improved by using a number of different techniques.

Thus, the content the users provide to the IR system can be analyzed (e.g. the queries they submit to a search engine). This way it is possible to optimize the results or to provide some feedback which the users can asses in their information retrieval process (e.g. a search engine could provide more focused results or query suggestions).

In the late 1990s seminal studies analyzing Web user's information retrieval process (e.g. \cite{Bateman1998} or  \cite{Silverstein1998}) reported important differences from users in other information retrieval scenarios. Those papers also included a categorization of queries using a predefined set of topics in an attempt of identifying the most interesting subjects for Internet users.   

Such analyses were performed in a supervised way once the retrieval process was over so no benefit could be extracted for the current user. Nowadays, however, there's a growing interest on performing this categorization in real time so the user can get a direct benefit. Google Trends\footnote{http://google.com/trends}, or Twitter trending topics can be seen as an automated version of those first studies allowing us to know what are the most popular subjects for the community. 

Another useful approach is extracting entities from the texts sent by the user. The underlying idea behind this method is the detection of groups of words, usually called \textit{entities}, which deserve a special treatment as they usually give name to a given topic or interest.

Traditionally, search engines give the user the ability to group keywords into phrases to stress a special relationship between them (e.g. \verb+"The hurt locker"+ \verb+oscars+). Such a feature, although very useful, has been reported to not be widely used (e.g. \cite{Spink2006} reports only 10-20\% of the queries using query operators) specially by inexperienced users who could benefit more from it (e.g. \cite{Hoscher2000} shows a significant difference between the average user and a team of expert Internet users). Hence, it could be extremely useful for users that the search engine detected such entities in an automatic way to better focus the search or to suggest the user a new query.

In fact, this very same technique was used by \cite{Kiseleva2010} to extract product brands and features (e.g. memory size, CPU speed...) in an e-commerce search engine. This way the user could get a better information retrieval experience by issuing more "natural" queries while still being "understood" by the system.

Unfortunately, query segmentation tend to be performed by means of statistical or machine learning methods which rely on a prior large corpus of training data. Nonetheless to say, such data is processed in advance during the training phase because of the unfeasibility of processing the corpus in real time when the queries are submitted.

Hence, performance is achieved at the cost of being unable to react to new types of queries (e.g. written on a different language or from a topic not appearing in the corpus). This can affect a considerable amount of the users of our IR system because a significant number of queries are submitted just once to the information retrieval system (\cite{Bateman1998}, \cite{Beitzel2007}, \cite{Saracevic2001a}) but, more importantly, this issue is critical in an ecosystem driving towards 'real time' where users demand fresh content faster than ever.

In this paper the authors propose and evaluate a new query segmentation approach designed to be as light and fast as possible; avoiding costly preprocessing or large datasets. The main goal is being able to create a system which can adapt to real time changes by just analyzing the information relevant for the current query.

\section{Previous Research}

Worries about the performance of NLP techniques applied to large datasets is neither new nor exclusive of Web Information Retrieval.

\cite{Zhai1997} argued that large document collections would need algorithms more efficient than those from previous IR research as the corpora began to grow. That author was specially interested in processing queries for entity extraction.

Nonetheless to say, noun phrase segmentation is not a novel area and large research efforts have been applied to this question (e.g. \cite{Gao2004}, \cite{Tan2008}, \cite{Bergsma2007} or \cite{Bendersky2009}).

Concerning Web Information Retrieval, noun phrase segmentation has been applied to query segmentation. However, it has been done in the same way it is applied to longer documents; that is, without taking into account the special features and requirements of search queries (e.g. they are shorter, much more schematic than a document, and users choose the vocabulary with different criteria than when writing a longer text).  

One possibility is to take advantage of the size and variety of available resources on the Web to improve not the algorithms but the dataset which should be analyzed. For instance, \cite{Risvik2003} applies some statistic processing over search engine query logs instead of using a traditional of corpus of web documents, assuming that users tend to segment queries in a reliable way. \cite{Tan2008} uses both a corpus of documents and information from Wikipedia to find query entities.

\section{Proposal Description}

\subsection{Characteristics Overview}

This authors are aware of the requirements of current web development. At this moment, an emerging ecosystem of real time applications is growing in importance, and well known algorithms are being adapted to this settings and new methods are being proposed.

In this new Web environment the context changes very quickly as the news spread in real time through web communities (e.g. Michael Jackson's death or Iran elections news arrived more quickly through Twitter's trend topics than through any news agency).

This changing web environment lead to the main requirement of our approach: being able to offer good results not only for unexpected queries, but also for queries which relate to drifting topics of interest.

As an example, the query \verb+icelandic volcano+ could have probably be segmented as \verb+[icelandic] [volcano]+ before April 2010; after the impact of the eruptions of Eyjafjallajökull on air travel a much better segmentation could probably be \verb+[icelandic volcano]+.

We are not saying, by no means, that methods based on prior training are not fast or feasible for this task. They are extremely fast after being trained and they have been applied to query segmentation. For instance, \cite{Tan2008} propose describe a method based on processing a large corpus in addition to Wikipedia titles, to pre-compute query segmentations for n-grams before any actual query was provided. Their method relies on result caching and, hence, they report a query segmentation rate of 500 queries per second.

It is the speed of topic drifting and how to handle that changing trends which we feel can't be properly achieved by using methods relying on training data and heavy result caching. Because of this, our approach consist of performing query segmentation on-the-fly. Of course, a balance should be reached as the results must be provided in a reasonable amount of time, in fact, repeated queries could and should be cached in order to avoid duplicate computations.

Authors are also conscious that returning fresh results demands a light algorithm which may not achieve as good performance as previous approaches based on training data. However, we feel that a certain lack of precision could be tolerable in exchange for a larger adaptability to drifting and evolving topics.

\subsection{Entities}

Previous researchers have discussed the nature of the MWUs a query segmentation algorithm should be able to found.

\cite{Tan2008} argues that, to be correct, entities must have a 'complete semantic meaning'.  This criteria introduces some ambiguity about the semantic completeness of an entity that is difficult to handle by any algorithm. Additionally, given the multilinguality nature of the Web, determining semantic completeness in an automatic fashion for any given entity in any given language could be insurmountably hard.

Other authors (e.g. \cite{Risvik2003} or \cite{Bergsma2007}) are not such concerned about the meaning of the queries and, thus, they assume that any set of words highly related according to the chosen statistical measure would be a MWU. This approach better handles some issues --such as multilinguality-- but it can also be noisy and nothing prevents the system of extracting collocations or idioms as MWUs (e.g. "good morning" or "birds of a feather").

In this paper we will refer as \textit{Entities} to those MWUs which the algorithm found to be statistically relevant.

\subsection{Snippet Sources}

Actual details about our method are provided below, at this moment, it is enough to say it is based on the use of snippets. Snippets are short extracts of web documents usually containing some keywords. They are commonly seen in result pages of search engines but we are open to also consider other sources of snippets (e.g. a Twitter message containing some of the keywords from the query could also be considered a snippet). However, for the purpose of this paper, we have only used snippets from three major search engines (Google\footnote{http://code.google.com/intl/es-ES/apis/ajaxsearch/documentation/\#fonje}, Bing\footnote{http://msdn.microsoft.com/en-us/library/dd251056.aspx} and Yahoo!\footnote{http://developer.yahoo.com/search/boss/}) exploiting their respective APIs. This way, we have been able to compare the performance of our method using different sources of snippets.

\subsection{Statistics Measures}

In addition to our own method we tested several statistical measures previously employed in MWU detection; namely, Mutual Information, SCP, Phi, Dice and LogLike. \cite{FerreiraDaSilva1999} devised a way to generalize those well-known measures from bigrams to n-grams of any arbitrary length and, hence, we have used those more general versions. However, due to space constraints we are not able to give any detail about them and, so, further details and formulas are to be found in the mentioned paper.

As we have already said we have devised another method based on the use of snippets. The underlying idea is really simple. First, all possible n-grams in the query are produced. Then, the frequency for each n-gram in the snippets is to be found and those n-grams below a certain threshold are discarded. Finally, the query is segmented by using the most frequent remaining n-grams. The aforementioned threshold limits the minimum number of snippets a n-gram should appear in; for instance, a 0.5 threshold means that those n-grams not appearing in 50

We tested four different values for this parameter (namely, 0.25, 0.5, 0.75 and 1) so we can study how it affects to the performance of our method.

\section{Evaluation}

\subsection{Datasets}

In order to evaluate the different query segmentation methods we needed some previously segmented queries. Those queries could have been segmented by "experts" or by actual search engine users.

A dataset segmented by experts should provide a consistent set of segmentations (e.g query \verb+new york travel guide+ would probably appear as \verb+[new york]+ \verb+[travel guides]+ in all its instances). However, according to \cite{Bergsma2007} the level of agreement between experts is relatively low (between 57.6\% and 60.8\%).

In contrast, a dataset segmented by actual search engine users should provide real information on how the users expect the search engine to group concepts; but with a much higher hetereogenity in the available segmentations because different users could understand concepts in different ways and, thus, would stress different segments within the queries.

Fortunately enough, we were able to found one dataset of each kind. The expert-segmented query log is the one described by \cite{Bergsma2007}. This dataset (from now on we will refer to it as aol-bergsma-wang-2007) is a subset of the AOL query log \cite{Torgeson2006} where the queries were manually segmented by the original authors of \cite{Bergsma2007}.

The second dataset, the one comprising queries segmented by actual users, was obtained from Dogpile SearchSpy. This second dataset (from now on searchspy-2010) is only comprised of queries including paired double-quotes which enclose two or more terms. 

In both datasets queries were lowercased and to obtain the snippets double-quotes were removed to avoid the search engine to exploit that information.

\subsection{Evaluation Methodology}

At a first glance, one could think of using the percentage of correctly segmented queries to measure the effectiveness of the different methods. However, this approach does not take in consideration the existence of different degrees of segmentation correctness. For instance, the query \verb+new york travel guides+ can be segmented in several different ways (\verb+[new york] [travel guides]+, \verb+[new york]+ \verb+travel guides+, \verb+new [york travel guides]+, etc). We could think of \verb+[new york]+ \verb+[travel guides]+ as the best segmentation and consider all the other ones as wrong segmentations. However, \verb+[new york]+ \verb+travel+ \verb+guides+ doesn't seem as wrong as \verb+new [york travel] guides+ or \verb+new york travel+ \verb+guides+.

In addition to these varying degrees of quality, it must be noticed that segmentation is a user-dependent task and two different users --even experts, as shown by \cite{Bergsma2007}-- can provide different segmentations; hence, a binary decision does not seem the best choice.

Both issues drive us to look for a different evaluation method judging individual segmentation decision instead of queries as a whole item.

We choose to use the well-known measures of precision and recall but taking into account not the queries but the blanks between segments both in the reference and the segmentation provided by each algorithm. This was based on the work of \cite{Makhoul1999} and is a pretty common approach when evaluating segmentation tasks. For instance, in the previous example query \verb+new york travel guides+ and assuming the correct segmentation is \verb+[new york] [travel guides]+, the segmentation \verb+[new york] travel guides+ have a precision of 1 and a recall of 0.5 while segmentation  \verb+new [york travel] guides+ have both 0 precision and 0 recall.

\subsection{Positive Bias from Snippets Sources}

\label{subsection:evaluation:bias}

As we have already said, all of the methods to be evaluated rely on snippets to produce results. For the study described in this paper the snippets are obtained from different major search engines. Modern Web search engines take into account the relative positions of keywords both in the queries and in the results and, thus, one could argue that segmentation performance is not actually a produce of the different evaluated methods but of the search engine employed. In order to measure the bias introduced by the search engine, and dispel any concern about the actual performance of the proposed technique, we run the algorithms in three different "flavors". Each flavor rearranged the queries in a different way before submitting it to the search engine to obtain the snippets.

The first one sends the queries after removing all the double-quotes; the second one sends the original queries, that is including double-quotes; and the third one not only removes the double-quotes but also reverses the order of the terms in order to remove any term collocation within the query that could be used as a "clue" by the search engine. For instance, the query \verb+"new york" travel guides+ would be submitted by each of the different flavors as: \verb+new york travel guides+, \verb+"new york" travel guides+\, and \verb+guides+ \verb+travel york new+.

This way, we have a base case in which queries are submitted with no segmentation information but keeping keywords ordering, a best case where all the segmentations within the query are preserved, and a worst case where neither segmentations nor the keywords ordering are preserved.

By doing this it could be possible to appreciate, albeit somewhat indirectly, the impact term collocation could exert on the results and, consequently, on the snippets and the different methods performance. Thus, if term collocations were heavily used by search engines to produce results then the differences between flavors 1 and 2 should be minimal. Additionally, by reversing the order of the terms in the query all valid collocations are removed and spurious ones are introduced; this should mangle the search engine results and, hence, make much harder to find valid segmentations. Of course, if even under this hard circumstances the methods manage to find correct segmentations we should discard the hypothesis of the search engine doing the work of detecting entities and should instead accept that the algorithms are actually finding them by means of the information available in the snippets.

\section{Experiment Results}

\subsection{Performance on \textit{searchspy-10} Query Log}

Table \ref{table:searchspy-results} shows the P, R, and F measures for each statistic measure and each snippet source used in the experiments.

\begin{table}[htb]

  \begin{tabularx}{\textwidth}{ | l | X | X | X | X | }

  \hline

  Measure & Snippet Source & P & R & F \\ \hline

  \multirow{3}{*} {Mutual Information} & Bing & 0.6834 & 0.5481 & 0.6158 \\  

  & Boss & 0.7464 & 0.6210 & 0.6837 \\  

  & Google & 0.5374 & 0.4349 & 0.4862 \\  \hline

  \multirow{3}{*} {SCP} & Bing & 0.6327 & 0.6216 & 0.6271  \\  

  & Boss & 0.6609 & 0.6481 & 0.6545 \\  

  & Google & 0.6145 & 0.6089 & 0.6117 \\  \hline

  \multirow{3}{*} {Phi} & Bing & 0.6888 & 0.5367 & 0.6128 \\  

  & Boss & 0.7530 & 0.6061 & 0.6795 \\  

  & Google & 0.5411 & 0.4246 & 0.4829 \\  \hline

  \multirow{3}{*} {Dice} & Bing & 0.6888 & 0.5354 & 0.6121 \\  

  & Boss & 0.7519 & 0.6057 & 0.6788 \\  

  & Google & 0.5405 & 0.4241 & 0.4823  \\  \hline

  \multirow{3}{*} {Loglike} & Bing & 0.7053 & 0.6383 & 0.6718 \\  

  & Boss & 0.7372 & 0.6663 & 0.7017 \\  

  & Google & 0.5349 & 0.4715 & 0.5032 \\  \hline

  \multirow{3}{*} {Entity Frequency (25)} & Bing & 0.7336 & 0.6446 & 0.6891 \\  

  & Boss & 0.8089 & \textbf{0.7375} & \textbf{0.7732} \\  

  & Google & 0.5873 & 0.5298 & 0.5585\\  \hline

  \multirow{3}{*} {Entity Frequency (50)} & Bing & 0.7447 & 0.5992 & 0.6719 \\  

  & Boss & 0.8238 & 0.6858 & 0.7548 \\  

  & Google & 0.5990 & 0.4938 & 0.5464 \\  \hline

  \multirow{3}{*} {Entity Frequency (75)} & Bing & 0.7510 & 0.5442 & 0.6476 \\  

  & Boss & 0.8304 & 0.6204 & 0.7254 \\  

  & Google & 0.6026 & 0.4450 & 0.5238 \\  \hline

  \multirow{3}{*} {Entity Frequency (100)} & Bing & 0.7530 & 0.5095 & 0.6312 \\  

  & Boss & \textbf{0.8329} & 0.5701 & 0.7015 \\  

  & Google & 0.6030 & 0.4084 & 0.5057 \\  \hline

\end{tabularx}

  \caption{Performance using different statistic measures and snippet sources for queries in \textit{searchspy-10} query log}

  \label{table:searchspy-results}

\end{table}

When Yahoo! Boss or Bing are used as snippet sources, F measures vary between 0.6 to 0.77; this is a reasonable performance for a technique that relies on relatively small information (10 snippets of text as a maximum) which would make statistical algorithms to work in an unreliable way.

Quite surprisingly, using Google as the snippet source drops F measures to very low values (from 0.48 to 0.61). This is shown in table \ref{table:searchspy-searchengine-results} where a difference of 0.15 is shown between Yahoo! Boss and Google.

\begin{table}[htb]

\begin{center}

\begin{tabular}{|l|c|c|c|}

\hline

 Snippet Source & P & R & F \\ \hline

 Bing & 0.7090 & 0.5753 & 0.6422 \\ \hline

 Boss & \textbf{0.7717} & \textbf{0.6401} & \textbf{0.7059} \\ \hline

 Google & 0.5734 & 0.4712 & 0.5223 \\ \hline

\end{tabular}

\end{center}

\caption{Average P, R and F measures for each snippet source for queries in \textit{searchspy-10} query log}

\label{table:searchspy-searchengine-results}

\end{table}

The main reason for such differences, beyond the quality of the returned results, is the amount of queries from the query log that Google couldn't find snippets for.

All of the snippet sources showed this lack of results for certain queries, but in the case of Google there were about 1,100 queries which didn't have any associated results. In the case of Bing less than 500 queries obtain no results, while Yahoo! did not return results for less than 300 queries.

We could have removed those "failed" queries but because the proposed system is highly dependent on the snippet source we feel it was important to measure performance using the whole query log.

Concerning to the statistical methods, the \textit{Entity Frequency} achieves the better performance, despite being the simplest one, with a difference of 0.07 from \textit{Loglike}, the second top-ranked algorithm.

\subsection{Performance on \textit{aol-bergsma-wang-2007} Query Log}

In table \ref{table:aol-results} we can see the results of our method when running on \textit{aol-bergsma-wang-2007} query log.

\begin{table}[htb]

  \begin{tabularx}{\textwidth}{ | l | X | X | X | X | }

  \hline

  Measure & Snippet Source & P & R & F \\ \hline

  \multirow{3}{*} {Mutual Information} & Bing & 0.7666 & 0.7240 & 0.7453 \\  

  & Boss & 0.7974 & 0.7557 & 0.7765 \\  

  & Google & 0.7833 & 0.7335 & 0.7584 \\  \hline

  \multirow{3}{*} {SCP} & Bing & 0.7084 & 0.7507 & 0.7295  \\  

  & Boss & 0.7234 & 0.7752 & 0.7493 \\  

  & Google & 0.6953 & 0.7512 & 0.7233 \\  \hline

  \multirow{3}{*} {Phi} & Bing & 0.8159 & 0.7362 & 0.7760 \\  

  & Boss & 0.8421 & 0.7655 & 0.8038 \\  

  & Google & 0.8269 & 0.7527 & 0.7898 \\  \hline

  \multirow{3}{*} {Dice} & Bing & 0.8182 & 0.7423 & 0.7802 \\  

  & Boss & 0.8387 & 0.7644 & 0.8016 \\  

  & Google & 0.8256 & 0.7545 & 0.7901 \\  \hline

  \multirow{3}{*} {Loglike} & Bing & 0.7842 & 0.8038 & 0.7940 \\  

  & Boss & 0.8012 & \textbf{0.8264} & \textbf{0.8138} \\  

  & Google & 0.8053 & 0.8238 & 0.8145 \\  \hline

  \multirow{3}{*} {Entity Frequency (25)} & Bing & 0.7183 & 0.7895 & 0.7539 \\  

  & Boss & 0.7374 & 0.8162 & 0.7768 \\  

  & Google &  0.7150 & 0.7886 & 0.7518\\  \hline

  \multirow{3}{*} {Entity Frequency (50)} & Bing & 0.7810 & 0.7218 & 0.7514 \\  

  & Boss & 0.8075 & 0.7433 & 0.7754 \\  

  & Google & 0.7751 & 0.7290 & 0.7521 \\  \hline

  \multirow{3}{*} {Entity Frequency (75)} & Bing & 0.8283 & 0.6230 & 0.7256 \\  

  & Boss & 0.8407 & 0.6507 & 0.7457 \\  

  & Google & 0.8340 & 0.6463 & 0.7402 \\  \hline

  \multirow{3}{*} {Entity Frequency (100)} & Bing & 0.8427 & 0.5430 & 0.6929 \\  

  & Boss & \textbf{0.8733} & 0.5590 & 0.7161 \\  

  & Google & 0.8505 & 0.5534 & 0.7019 \\  \hline

\end{tabularx}

  \caption{Performance using different statistic measures and snippet sources for queries in \textit{aol-bergsma-wang-2007} query log}

  \label{table:aol-results}

\end{table}

For this query log, F measures achieves better performance (between 0.07 for Boss and 0.23 for Google) as we can see in table \ref{table:aol-searchengine-results}. The main reason for this better performance is that very few queries obtained no results in the snippet sources because \cite{Bergsma2007} selected only those queries with at least one clicked result in the AOL query log.

If those queries without results are removed the differences between snippet sources are not significant, but, again, using Yahoo! Boss achieves the better performance.

\begin{table}[htb]

\begin{center}

\begin{tabular}{|l|c|c|c|}

\hline

 Snippet Source & P & R & F \\ \hline

 Bing & 0.7848 & 0.7149 & 0.7499 \\ \hline

 Boss & \textbf{0.8069} & \textbf{0.7396} & \textbf{0.7732} \\ \hline

 Google & 0.7901 & 0.7259 & 0.7580 \\ \hline

\end{tabular}

\end{center}

\caption{Average P, R and F measures for each snippet source for queries in \textit{aol-bergsma-wang-2007} query log}

\label{table:aol-searchengine-results}

\end{table}

Concerning the algorihms, \textit{Entity Frequency} behaves better with this query log than with the previous one but a more important boost is the one achieved by other statistic measures such as \textit{Dice}, \textit{Mutual Information} or \textit{Loglike}. In fact, \textit{Loglike} is the algorithm which achieves the highest values for R and F measures, although \textit{Entity Frequency} still shows the highest value for Precision.

\subsection{Bias Results}

As we have previously explained, it should be confirmed that the results are not due to the search engines performing query entity extraction and thus returning results for those entities.

Due to space constraints tables summarizing the results for what we have called best and worst flavors can't be shown, but Table \ref{table:variation-bias-results} shows the variation obtained in F measure for the best and worst cases as described in Section \ref{subsection:evaluation:bias}.

It can be noticed that forcing quotes in the queries led to an increase in the number of queries no obtaining results.As we have seen before, this lack of results introduces some bias on the snippets source performance so we are going to clean up the datasets of those queries with no results This way, variations on the snippets source performance will not be caused by the quality of the query but by the snippet source itself.

\begin{table}[htb]

\begin{center}

\begin{tabular}{|l|c|c|c|c|}

\hline

 Snippet Source & \multicolumn{2}{|c|}{\textit{searchspy-10}} & \multicolumn{2}{|c|}{\textit{aol-bergsma-wang-2007}} \\ \hline

  & Quoted & Reversed & Quoted & Reversed \\ \hline

 Bing & 0.0902 & -0.0029 & 0.0785  & -0.0013 \\ \hline

 Boss & 0.0645 & -0.0160 & 0.0443  & -0.0149 \\ \hline

 Google & 0.0742 & -0.0124 & 0.0580  & -0.0142 \\ \hline

\end{tabular}

\end{center}

\caption{Average variations in F measures for best (quoted) and worst (reversed) scenarios}

\label{table:variation-bias-results}

\end{table}

Although differences in best cases are not really significant (between 0.04 and 0.09) when we compare them against differences with worst cases we find that best cases are very different from the base case.

So, in short, by means of these experiments comparing best and worst case scenarios, we can safely assert that (1) our technique is actually performing query segmentation --certainly by means of the snippets-- and (2) it is not inadvertently taking advantage of any underlying query segmentation or entity detection phase performed by the search engine.

\section{Conclussions}

\subsection{Summary}

In this paper we have revisited the challenges IR systems must face with regards to the lack of context for user goals. Such challenges are growing because of the currently emerging real-time scenario which requires new solutions to handle not only such a lack of context, but also the fast drift of the topics underlying a given query.

We have described how query entity extraction can be a useful tool to tackle with this problem, and we have proposed using short snippets to obtain --by means of simple statistical methods-- the most relevant MWUs.

Such methods have been implemented and some experiments have been performed on two previously segmented query logs: one of them by a panel of experts and the other one by actual Web searchers. Preliminary results are pretty good (F-measure is about 0.8) and, thus, further research in this line should be done.

\subsection{Snippet Source Dependency}

\textit{searchspy-10} query log reflects real users' behavior with regards to query segmentation (let's remember it comprises double-quoted queries as issued by the users). The main conclusion one can reach from the experiments conducted on this query log is that performance is highly dependent on the underlying snippet source.

Preliminary work using the Wikipedia search engine as a source of snippets has shown that Wikipedia can be a very good source for certain kind of queries but not that good for other ones.

\subsection{Statistic algorithm election}

Two algorithms outperform the others in both experiments: Loglike and Entity Frequency. Hence, it seems that final choice for our method is between these two statistical approaches.

On one hand, Loglike achieves the highest performance for all the experiments, its behavior is highly irregular and does not handle very well queries from "actual" users (i.e. those from the \textit{searchspy-10} query log). On the other hand, Entity Frequency provides much more consistent results for both query logs achieving similar performance.

Besides, Entity Frequency has been studied with different thresholds, exhibiting different performance values --specially for precision and recall-- and this is also a variable to take into account.

\section{Future Work}

We have described a preliminary implementation which achieves good results; nevertheless, some questions remain open and they deserve further research.

\paragraph{Real Time Snippets}

More work must be done on integrating different snippet sources directly related to real-time Web, such as micro-blogging and blogging streams (e.g. Twitter, or Google Blog Search), or collaborative sites covering different topics (e.g Wikipedia).

\paragraph{Snippet Sources Integration}

The use of various complementary sources of snippets could improve the performance provided a feasible way to select the most appropriate source or to blend all the results is available.

\section{Acknowledgments}

This work was partially financed by grant UNOV-09-RENOV-MB-2 from the University of Oviedo

\bibliographystyle{plain}

\bibliography{paper_ceri}

\end{document}